\documentclass[preprint,aps,superscriptaddress]{revtex4}
\usepackage{graphicx,epsfig,amssymb,amsmath,array}
\usepackage{relsize,placeins,float}
\usepackage[usenames]{color}

\begin{document}
\title{Phase control of escapes in the fractional damped Helmholtz oscillator}
\author{Mattia Coccolo}
\affiliation{Nonlinear Dynamics, Chaos and Complex Systems Group, Departamento de F\'{i}sica,
Universidad Rey Juan Carlos, Tulip\'{a}n s/n, 28933 M\'{o}stoles, Madrid, Spain}
\author{Jes\'{u}s M. Seoane}
\affiliation{Nonlinear Dynamics, Chaos and Complex Systems Group, Departamento de F\'{i}sica,
Universidad Rey Juan Carlos, Tulip\'{a}n s/n, 28933 M\'{o}stoles, Madrid, Spain}
\author{Stefano Lenci}
\affiliation{Department of Civil and Building Engineering, and Architecture, Polytechnic University of Marche, 60131 Ancona, Italy}
\author{Miguel A.F. Sanju\'{a}n}
\affiliation{Nonlinear Dynamics, Chaos and Complex Systems Group, Departamento de F\'{i}sica,
Universidad Rey Juan Carlos, Tulip\'{a}n s/n, 28933 M\'{o}stoles, Madrid, Spain}
\date{\today}

\begin{abstract}
We analyze the nonlinear Helmholtz oscillator in the presence of fractional damping, a characteristic feature in several physical situations. In our specific scenario, as well as in the non-fractional case, for large enough excitation amplitudes, all initial conditions are escaping from the potential well. To address this, we incorporate the phase control technique into a parametric term, a feature commonly encountered in real-world situations. In the non-fractional case it has been shown that, a phase difference of $\phi_{OPT} \approx \pi$, is the optimal value to avoid the escapes of the particles from the potential well. Here, our investigation focuses on understanding when particles escape, considering both the phase difference $\phi$ and the fractional parameter $\alpha$ as control parameters. Our findings unveil the robustness of phase control, as evidenced by the consistent oscillation of the optimal $\phi$ value around its non-fractional counterpart when varying the fractional parameter. Additionally, our results underscore the pivotal role of the fractional parameter in governing the proportion of bounded particles, even when utilizing the optimal phase.

\end{abstract}

\maketitle

\section{Introduction}\label{sec:introduction}

Phase control is a crucial concept for complex dynamical systems. Nonlinear oscillators exhibit behaviors that defy simple periodicity, often showcasing rich and intricate dynamics. The control of their dynamical behaviors and, therefore, obtaining a desirable situation or avoiding an undesirable situation, is a relevant topic in the context of Nonlinear Dynamics. There are different ways of applying a control on a given dynamical system, that typically can be classified in two main types depending on how they interact with the system: feedback and non-feedback methods~\cite{BOCCALETTI:PR}. Among the non-feedback  methods, the phase control technique~\cite{Zambrano:2006,Seoane} is a typical scheme for controlling driven dynamical systems. Nonlinear oscillators described by a second-order differential equation with a fixed damping parameter, a nonlinear potential, and driven by an external periodic forcing are well-suited for the application of this method.

In such systems, phase control becomes a fundamental technique to analyze and tame oscillations. It involves defining and tracking the phase of the oscillator relative to a reference point, which is typically expressed as an angle or dimensionless quantity. This phase information allows researchers and engineers to understand the temporal relationships within the oscillator's dynamics, making it possible to predict, control, and synchronize its behavior. Phase control has applications in diverse fields, from physics and biology to engineering and information technology, offering insights into phenomena ranging from cardiac rhythms and brain activity to synchronized lasers and chaos-based secure communications. By harnessing the concept of phase control, we can better understand the complex, nonlinear dynamics of oscillatory systems, enabling advancements in science and technology.
A notable example of the application of phase control can be found in the field of neuroscience. When studying brain rhythms, researchers have employed phase control techniques to investigate neural synchronization and communication. For instance, Buzsaki and Draguhn \cite{Buzsaki}  demonstrated the significance of phase relationships in the coordination of neural ensembles during cognitive processes, shedding light on the brain's dynamic functioning. In the realm of physics, phase control is paramount for manipulating the synchronization of lasers, as exemplified by the work of Strogatz and Mirollo \cite{Strogatz1}. They discussed the concept of "phase-locking" in coupled oscillators, which is fundamental in laser physics, where the synchronization of laser light waves is essential for applications in communication and precision measurement. Furthermore, the study of heart rhythms and cardiac arrhythmias relies heavily on phase control techniques. Recent research \cite{Namasivayam,Monga} explored the use of phase control to investigate the onset of dangerous cardiac arrhythmias, providing insights into potential therapies for heart-related disorders. The application of this method has also been utilized to control the spread of epidemics \cite{DUARTE:2021}. All this shows that it remains a topic of active research with promising implications for diverse scientific and technological advancements.

In general, all these problems have been faced in the context of non-fractional derivatives. However, recent works show that, in several physical situations, the damping term can be fractional and it creates an important influence in the dynamical behavior of the system \cite{Coccolo_fr}. In this specific case, a fractional damping term introduce non-integer-order derivatives to the equations, allowing for the description of systems with memory effects, anomalous diffusion, or other non-standard behaviors. These fractional systems are of interest in various scientific fields, including physics, engineering, and applied mathematics \cite{Boroviec,Dafermos,Lu,Chellaboina,De,Elliott,Horr,Ding,Li,Strogatz}.

Here, we study the phase control technique in the nonlinear Helhmoltz oscillator with fractional damping. In Ref.~\cite{Seoane}, the authors studied and analyzed the phase control in the non-fractional Helmholtz oscillator. In this work, the authors show the optimal value of the phase for which most trajectories keep inside the well and they do not escape. Now, in our current situation, the dynamics of a Helmholtz oscillator with fractional damping can exhibit intriguing and rich behavior \cite{Coccolo_fr} since the fractional parameter $\alpha$ can also play the role of a control parameter and therefore is relevant for the escaping dynamics. The fractional damping term influences the rate at which energy is dissipated from the system, and the fractional order determines the memory effect or the long-term correlation in the damping process.  These features of the fractional derivatives can affect the phase control. The potential of the Helmholtz oscillator is considered a prototype for transient chaos and escape phenomena. Therefore, we extend what have been showed in~\cite{Seoane} and study the effect of the fractional parameter and the phase control on the escape of the particles. In particular, we analyze the attractor and the escape time of the particle, i.e., the final state of the trajectories, inside or outside the potential well, and the time to reach such state, respectively.

Our objective is to effectively manage the impact of minor disturbances as a control. In particular, we assume that the strength of our control is smaller than the external forcing. As a consequence, apart from optimizing energy consumption for controlling the trajectory, it shows a relevant characteristic of a control technique.

This paper is organized as follows. In Sec.~\ref{sec:Model Description}, we present the Helmholtz oscillator with the fractional damping. The phase control scheme and the consequences of the dynamical behavior of the system are presented in Sec.~\ref{sec:phasecontrol}. The effects of the fractional parameter when we vary the phase and the amplitude of the control forcing are described in Sec.~\ref{sec:phi}. Likewise, Sec.~\ref{sec:alpha} shows how the fractional parameter affects the control of escapes from the potential well for different initial conditions. Finally, the main conclusions and a discussion of the results are presented in Sec.~\ref{sec:conclusions}.

\section{Model description}\label{sec:Model Description}

 \begin{figure}[htbp]
  \centering
   \includegraphics[width=10.0cm,clip=true]{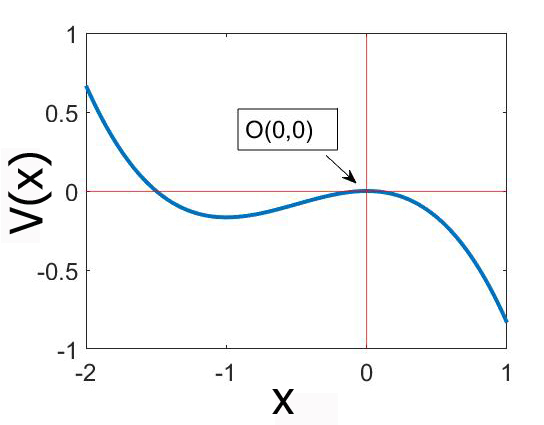}
   \caption{This plot represents the potential of the Helmholtz oscillator, the effect of the modulation term $[1+\epsilon\cos{(t+\phi)}]$ is to make the bottom of the well oscillate vertically.}
\label{fig:1}
\end{figure}

The Helmholtz oscillator is represented by a nonlinear second order differential equation considering the presence of the potential defined as $V(x) = -\frac{1}{2}x^{2} - \frac{1}{3}x^{3}$, as shown in
Fig.~\ref{fig:1}. It represents the equation of motion of a unit mass particle in a cubic potential under the influence of both a periodic forcing and a dissipative force and is given by equation:

\begin{eqnarray}\label{eq:classical_helmholtz}
\ddot{x} + \mu \dot{x} - x - x^2 =  F\cos(\omega t),
\end{eqnarray}
where $\mu$ is the damping parameter, $F$ and $\omega$ the forcing amplitude, and the forcing frequency, respectively, taking all of them positive values. This represents a simple paradigmatic example of a dynamical system with escapes.  In this system, normally, the damping term, considered like the first derivative in Eq.~(\ref{eq:classical_helmholtz}), has an impact proportional to the constant $\mu$ in the system. For our numerical simulations we fix $\mu=0.1$.

In what follows, the first derivative of Eq.~(\ref{eq:classical_helmholtz}) is replaced by an
order $\alpha$ fractional derivative, i.e, $\dot{x}(t)\rightarrow
d^{\alpha}x(t)/dx^{\alpha}$, for which we have the Helmholtz
Oscillator with a fractional damping term $\alpha$. This reads as follows:

\begin{eqnarray}\label{eq:fractional_helmholtz}
\ddot{x} + \mu \frac{d^{\alpha}x}{dx^{\alpha}} - x - x^2 =  F\cos(\omega t).
\end{eqnarray}
Adding a fractional damping term to the Helmholtz oscillator equation involves replacing the integer-order damping term with a fractional derivative term.  Mathematically, if $v(t) = \frac{d}{dt}x(t)$ is the velocity, then the fractional derivative of order $\alpha$ could be represented as  $v^\alpha(t) = \frac{d^{\alpha}x(t)}{dt^{\alpha}}$. This fractional derivative does not have a straightforward physical interpretation like the first-order derivative does, because it blends the effects of all past states of the system into the current state. In fact, if the first-order derivative of space with respect to time represents velocity, then a fractional derivative of space with respect to time can be thought of as a “fractional velocity,” which incorporates the memory effect of the velocity. This means that the fractional velocity at a given moment is influenced not only by the current rate of change of position but also by the entire history of the object’s motion. In practical terms, the fractional derivative introduces memory effects or non-local influences into the damping mechanism. Physically, the fractional damping term could represent phenomena such as viscoelasticity in materials, where the damping force depends not only on the current velocity of the oscillator but also on its past velocities. This means that the velocity has a history-dependent nature. It could also model systems with complex interactions between different parts, where the damping effect is influenced by the history of the motion \cite{Suzuki}. 

To integrate the system, Eq.~\ref{eq:fractional_helmholtz} can be written as a set of three fractional differential equations as follows:

\begin{eqnarray}\label{eq:system_helmholtz}
\frac{d^{\alpha}x}{dt^{\alpha}}  & =  & y \\ \nonumber \frac{d^{1-\alpha}y}{dy^{1-\alpha}}  & =  & z \\
\nonumber \frac{dz}{dt}  & = &  F\cos(\omega t) + x^2 + x -\mu y, \nonumber
\end{eqnarray}
where $z$ is a mathematical component coming from the transformation of the model into a fractional order system. To obtain the solution of Eqs.~(\ref{eq:system_helmholtz}), we use the {\it Gr\"unwald-Letnikov}~\cite{ALGORITHM:1} fractional derivative, for which the algorithm to numerically solve this system is given by

\begin{eqnarray}\label{eq:numerical_system_helmholtz}
x(t_{k})  & =  & y(t_{k-1})h^{\alpha} -
\sum_{j=\upsilon}^{k}c_{j}^{(\alpha)}x(t_{k-j}) \\ \nonumber
y(t_{k})  & =  & z(t_{k-1})h^{1-\alpha}  -
\sum_{j=\upsilon}^{k}c_{j}^{(1-\alpha)}y(t_{k-j})\\ \nonumber
z(t_{k})  & = &  \Psi h -
\sum_{j=\upsilon}^{k}c_{j}^{(1)}z(t_{k-j}), \nonumber
\end{eqnarray}
where $\Psi =  F\cos(\omega t_{k}) - x^2(t_{k}) - x(t_{k})-\mu
y(t_{k})$ and $h$ is the discrete-time step. The coefficients
$c_{j}^{\alpha}$ are the binomial coefficients derived in the
numerical scheme implemented, $c_{0}^{\alpha}=1$ and

\begin{eqnarray}
c_{j}^{\alpha}=(1-\frac{\alpha + 1}{j})c_{j-1}^{\alpha}.
\end{eqnarray}

The model equation that we will analyze once the phase control is introduced parametrically is the following

\begin{eqnarray}\label{eq:fractional_helmholtz_phase}
\ddot{x} + \mu \frac{d^{\alpha}x}{dx^{\alpha}} - x - [1+\epsilon\cos{(t+\phi)}] x^2 =  F\cos(\omega t).
\end{eqnarray}

The $\epsilon$ parameter is the modulation amplitude and the $\phi$ parameter is the phase. So, the potential becomes:
\begin{equation}
    V(x,t) = -\frac{1}{2}x^{2} - \frac{1}{3}[1+\epsilon\cos{(t+\phi)}]x^{3}.
\end{equation}
If we fix $\epsilon$ and $\phi$ the bottom of the potential well oscillate in function of $t$ periodically in the vertical direction.
This control implementation was used, in the non-fractional case, in Refs.~\cite{Zambrano:2006, Seoane} and it was also tested in an electronic circuit which mimics the dynamics of the system.

 \begin{figure}[htbp]
  \centering
   \includegraphics[width=16.0cm,clip=true]{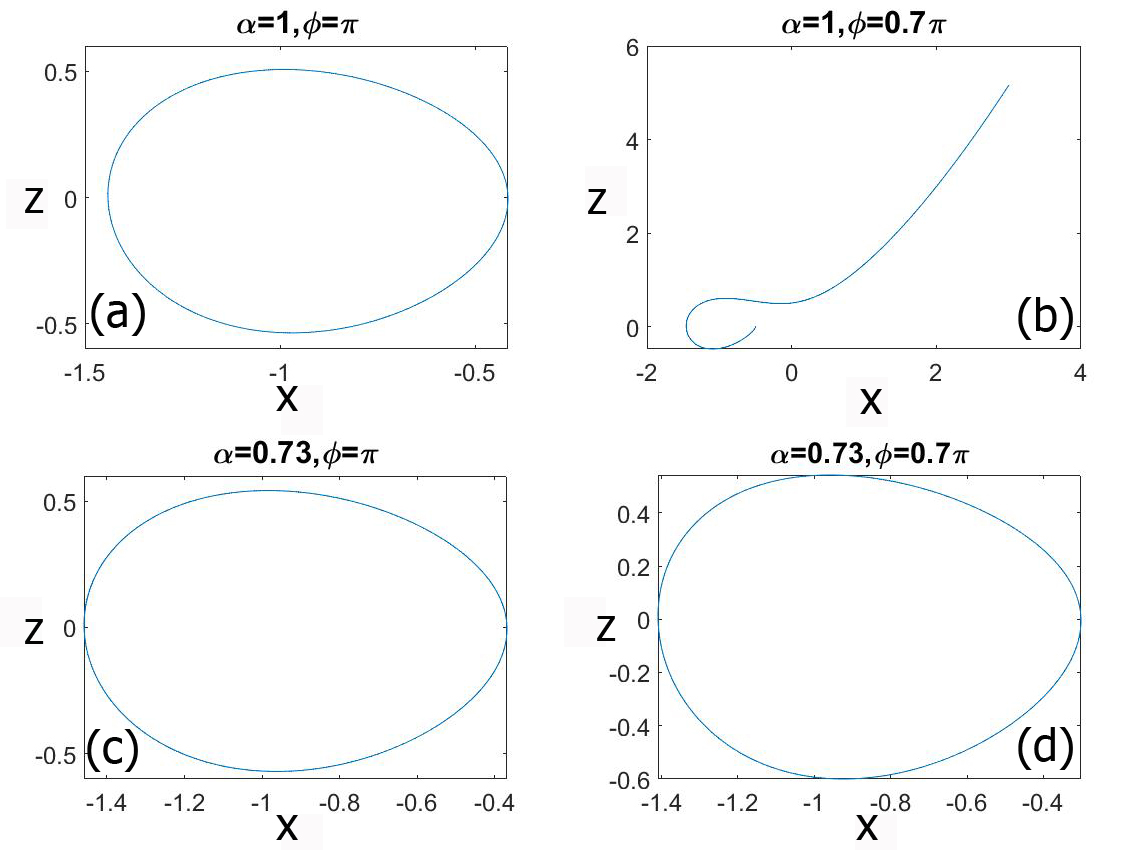}
   \caption{Here, we show some trajectories of the Helmholtz oscillator. The effect of the interaction between parameter $\alpha$ and $\phi$ is evident in panels (b) and (d). The initial conditions are $(x_0,\dot{x}_0)=(-0.5,-0.1)$ and $F=0.2$.}
\label{fig:2}
\end{figure}

\begin{figure}[htbp]
  \centering
   \includegraphics[width=8.0cm,clip=true]{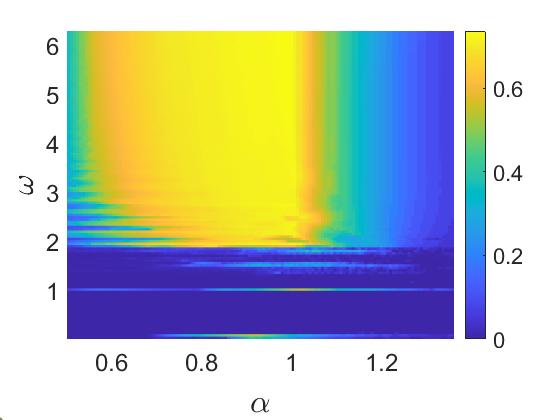}
   \caption{scape parameter set $(\omega-\alpha)$  for $F=0.2$, $\epsilon=0.1$ and $\phi=\pi$. The color bar shows the ratio over $1$ of the trajectories that do not leave the potential well by varying in every point the initial conditions in the region $[x_0]\times [y_0]=[-1.5,0]\times[-0.7,0.7]$. We can observe that, in our case and for our choice of parameters values, in which we have taken resonant frequencies,  $\omega= 1$ is an optimal value to avoid escapes as already stated in \cite{Seoane}.}
\label{fig:3a}
\end{figure}

\section{Phase control and the influence of the fractional derivative parameter}\label{sec:phasecontrol}

Here, we implement the phase control scheme in the fractional Helmholtz oscillator when  all initial conditions escape from the potential well. Also, we work with resonance frequencies between the periodic forcing and the modulation term. Therefore, we fix $F=0.2$, $\mu=0.1$ and $\omega=1$.  This choice of the frequency value takes into account resonant frequencies between the main forcing $F \cos \omega t$ and the control term $[1+\epsilon \cos(\omega t + \phi)]$. The value $\omega=1$ has been identified as optimal for preventing escapes from the well, as observed in the non-fractional case \cite{Seoane}. Nevertheless, we have computed Fig.~\ref{fig:3a}, wherein we demonstrate that, in the fractional case and for different values of the forcing frequency, $\omega=1$ locally yields the best performance in retaining most of the particles within the well. Therefore, the equation of motion reads:

\begin{eqnarray}\label{eq:fractional_helmholtz_phase_2}
\ddot{x} + 0.1 \frac{d^{\alpha}x}{dx^{\alpha}} - x - [1+\epsilon\cos{(t+\phi)}] x^2 =  0.2\cos t.
\end{eqnarray}

\begin{figure}[htbp]
  \centering
   \includegraphics[width=14.0cm,clip=true]{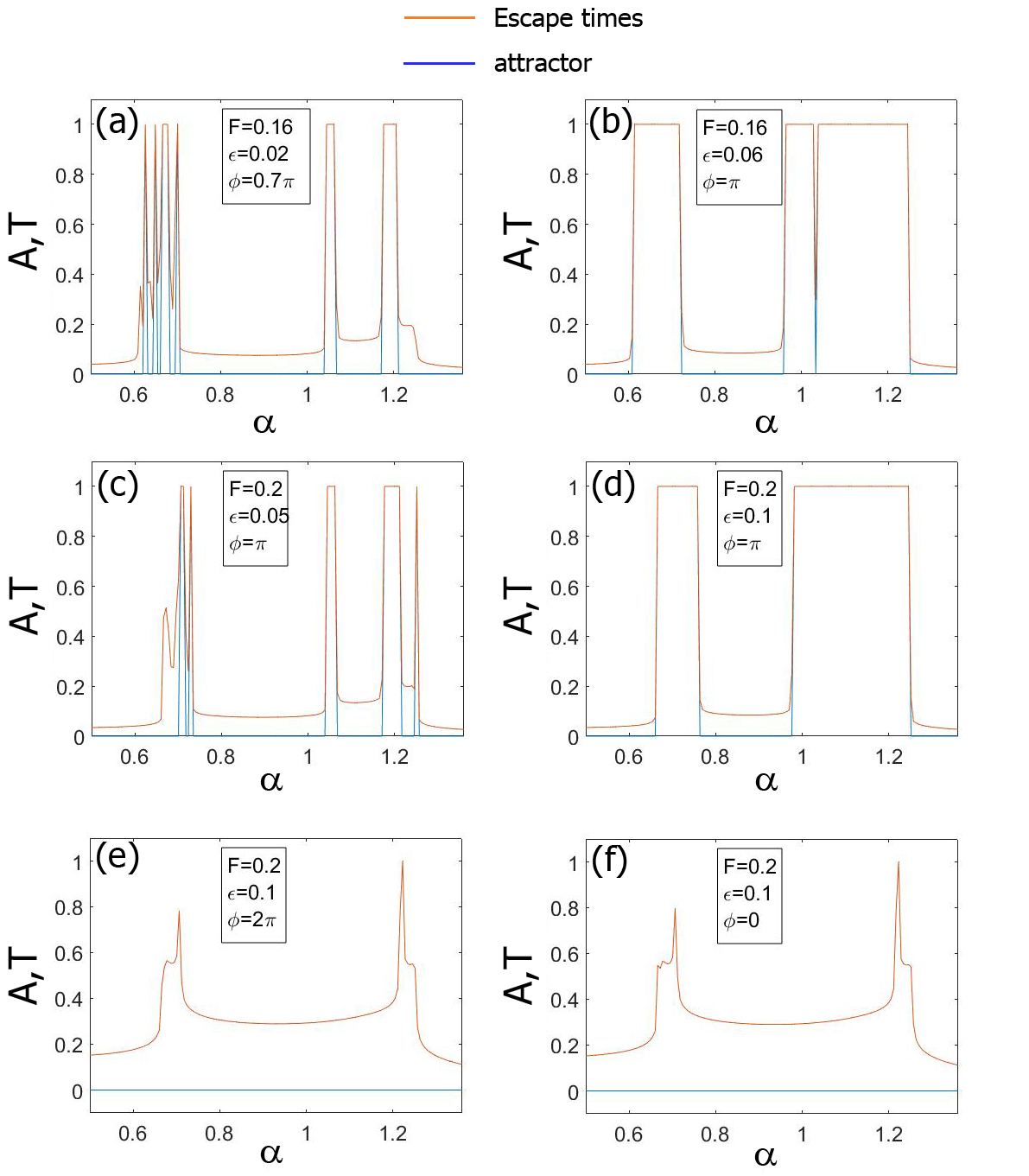}
   \caption{The attractive region $A$ (denoted by the blue line) and the escape times distribution $T$ (red line, it is normalized at the maximum $1$ instead of $125$ for a better comparison with the attractor) as a function of the fractional parameter $\alpha$ are shown in the panels (a-f). When the attractive region is equal to $1$ the particle does not escape, while when it is $0$ the particle escapes.  We can see that depending on the choice of the parameters $F,\epsilon$ and $\phi$, the value of $\alpha$ plays a relevant role to control the trajectories or to obtain different escape times.}
\label{fig:3}
\end{figure}

After switching on the phase control, that means setting the parameter $\epsilon\neq0$, we plot trajectories for different values of the parameters $\alpha$ and $\phi$ in Fig.~\ref{fig:2}.  Here, we  fix $\epsilon=0.1$ and initial conditions $(x_0,\dot{x}_0)=(-0.5,-0.1)$ and discard the transient dynamics.  As we can see in Fig.~\ref{fig:2}(a) the particle is controlled for $\phi=\pi$ in the non-fractional case ($\alpha=1$) and in the fractional case $\alpha=0.73$ as shown in Fig.~\ref{fig:2}(c). However, when we change to $\phi=0.7\pi$ the non-fractional case trajectory escapes from the potential well in Fig.~\ref{fig:2}(b), while in the fractional case it is trapped in it (see Fig.~\ref{fig:2}(d)). So the fractional parameter plays a major role to influence the escape or the control of the particle inside the well. For a better understanding of the effect of the fractional parameter, we show in Fig.~\ref{fig:3} an attractive region ($A$) and  the escape times ($T$) distribution as a function of $\alpha$ for different choice of parameters $F,\epsilon$ and $\phi$. The attractive region, denoted by the blue line, assumes the value $1$, when the particle is controlled inside the well and its escape time is infinite, or $0$ when it escapes. The escape times, denoted by the red line, have been normalized to $1$ instead of the maximum time of integration $t=125$ to compare them with the attractive region in the same conditions. We can see in Fig.~\ref{fig:3}(a-d) that the escape time and the attractive region follow similar patterns. Indeed, the information that we extract from the two curves is different. The blue curve $A$ tells us the value of the parameter $\alpha$ that control the particle inside the well. The orange curve $T$ indicates the impact of the parameter on the transient dynamics of the system. In fact, in Fig.~\ref{fig:3}(e) and Fig.~\ref{fig:3}(f), there are no controlled particles ($A=0$ for all $\alpha$ values), but the escape times change by varying the $\alpha$ parameter.  Moreover, our numerical experiments have shown that the particles can be controlled with a smaller value of $\epsilon$ when $F=0.16$. This is the reason why we have decided to keep on our analysis only with the case $F=0.2$ for which the particles always escape in the non-fractional case ($\alpha=1$) except for $\epsilon\approx0.1$ and $\phi \approx \pi$.

 \begin{figure}[htbp]
  \centering
   \includegraphics[width=16.0cm,clip=true]{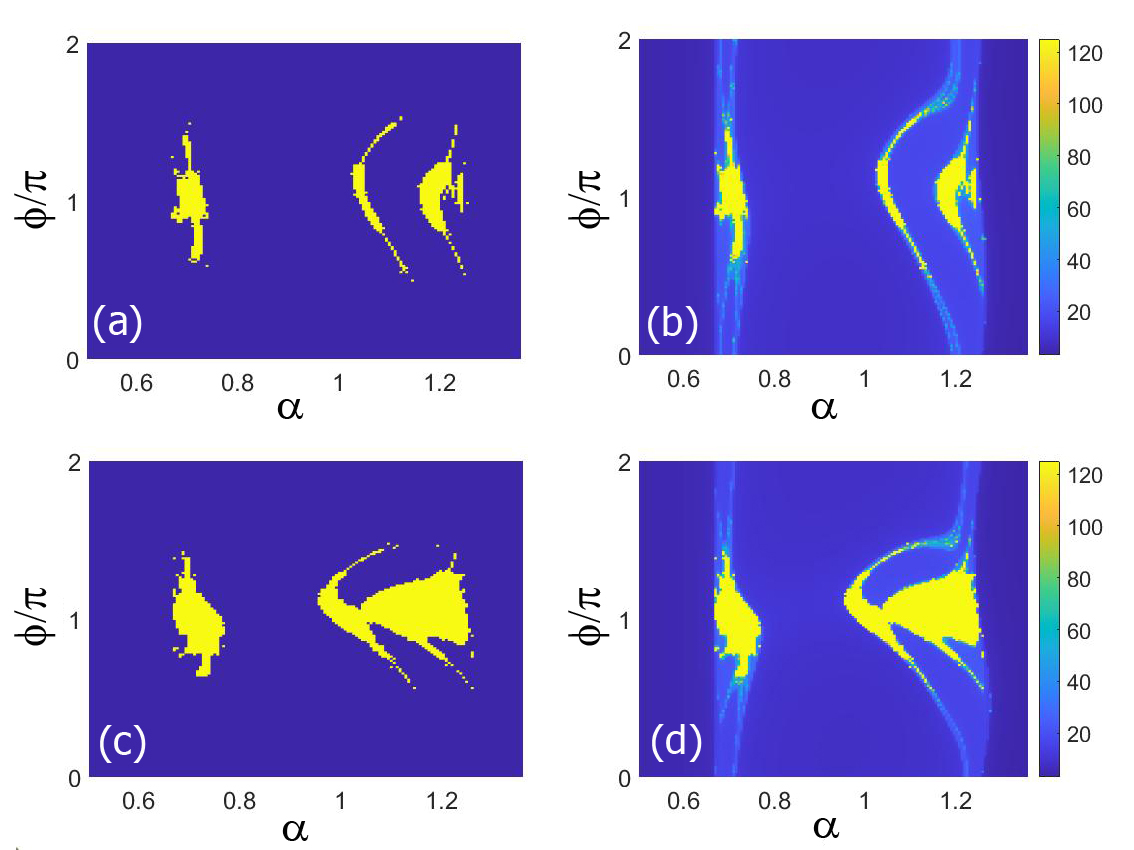}
   \caption{The attractive region, in the left column, and the escape times distribution, in the right column, in the parameter set $(\phi-\alpha)$,  for $F=0.2$ and $\epsilon=0.06$ (a) and (b),  and for $F=0.2$ and $\epsilon=0.1$ (c) and (d). The color code denotes the escape time of every trajectory. The yellow structures indicate the regions for which the particles spend more time to escape from the potential well.}
\label{fig:4}
\end{figure}

\begin{figure}[htbp]
  \centering
   \includegraphics[width=16.0cm,clip=true]{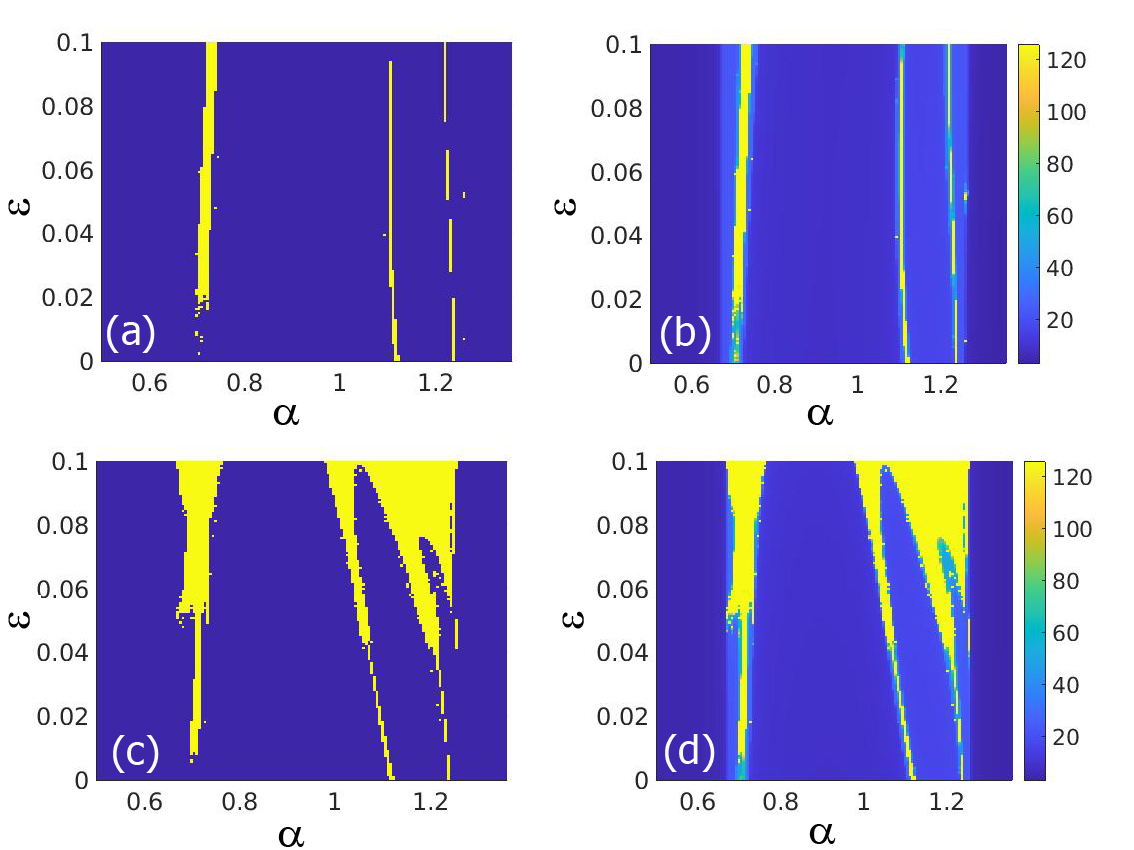}
   \caption{The attractive region, in the left column, and the escape times distribution, in the right column, in the parameter set $(\epsilon-\alpha)$,  for $F=0.2$ and $\phi=0.7\pi$  (a) and (b), and for $F=0.2$ and $\phi=\pi$  (c) and (d). The color code denotes the escape time of every trajectory. The yellow structures indicate the regions for which the particles spend more time to escape from the potential well.}
\label{fig:5}
\end{figure}

\begin{figure}[htbp]
  \centering
   \includegraphics[width=16.0cm,clip=true]{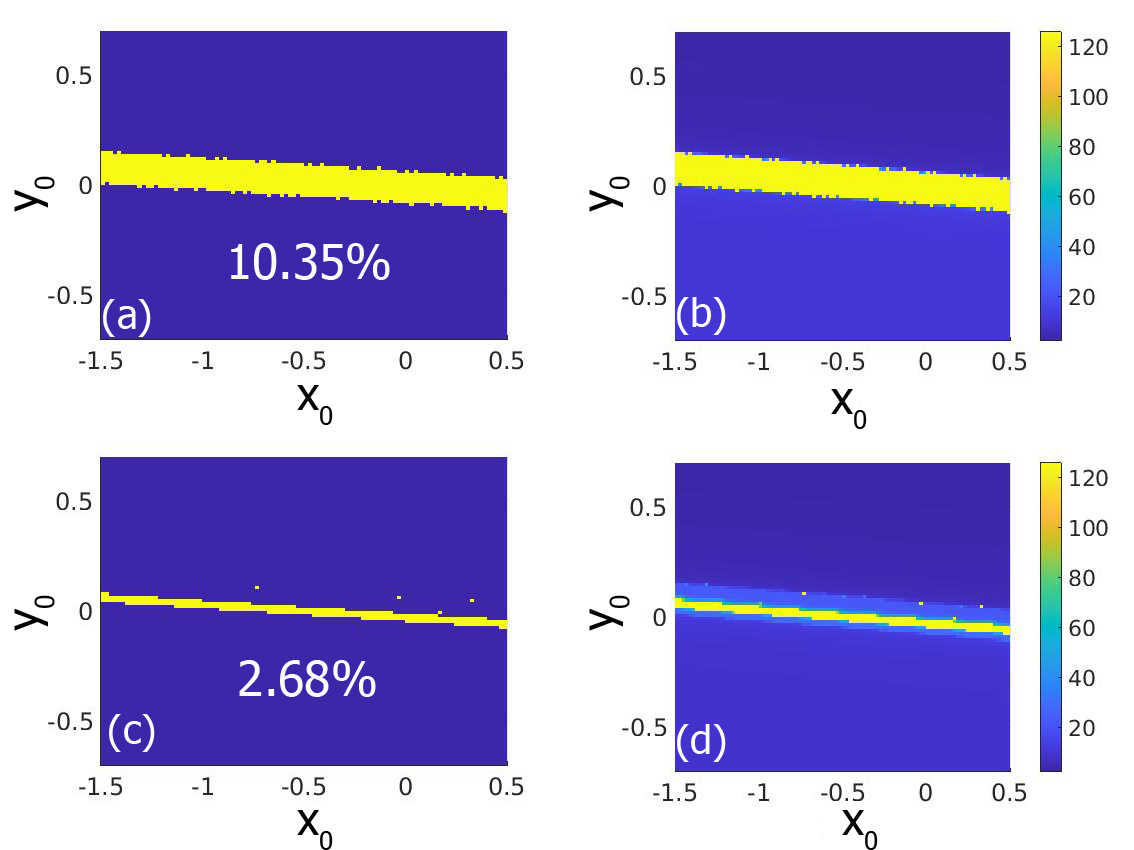}
   \caption{The basins of attraction, in the left column, and the escape times distribution, in the right column, in the initial conditions space $(x_0,y_0)$,  for $F=0.2$ and $\phi=\pi$  (a) and (b), and for $F=0.2$ and $\phi=0.7\pi$  (c) and (d). In all the panels $\epsilon=0.1, \alpha=0.73$ and the initial conditions are set in the region $[-1.5,0]\times[-0.7,0.7]$. The percentages inside the figures of the left column represent the ratio of the initial conditions for which the trajectories do not escape from the potential well.}
\label{fig:6}
\end{figure}

\begin{figure}[htbp]
  \centering
   \includegraphics[width=16.0cm,clip=true]{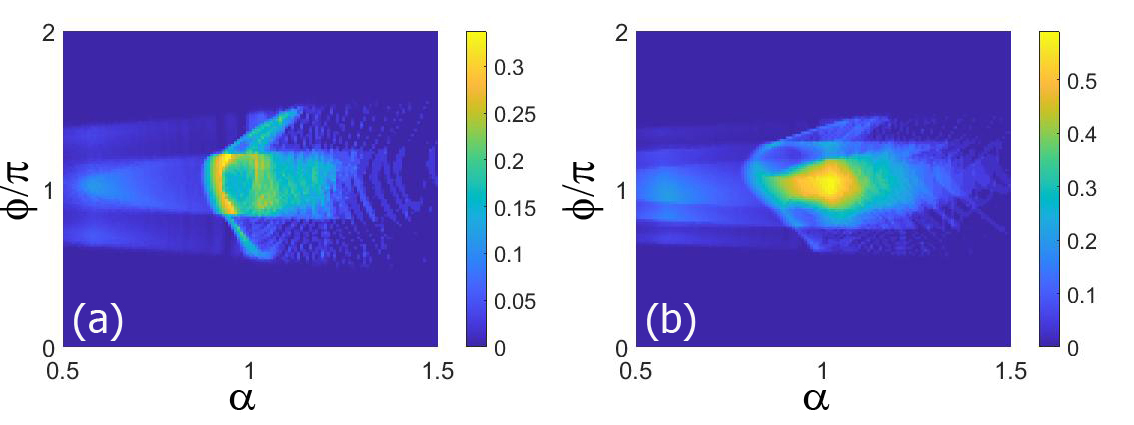}
   \caption{Escape parameter set $(\phi-\alpha)$,  for $F=0.2$ and $\epsilon=0.06$  (a) and $\epsilon=0.1$ (b). In all panels, the color bar shows the ratio over $1$ of the trajectories that do not leave the potential well by varying in every point the initial conditions in the region $[x_0]\times [y_0]=[-1.5,0]\times[-0.7,0.7]$.}
\label{fig:7}
\end{figure}
\FloatBarrier

\begin{figure}[htbp]
  \centering
   \includegraphics[width=16.0cm,clip=true]{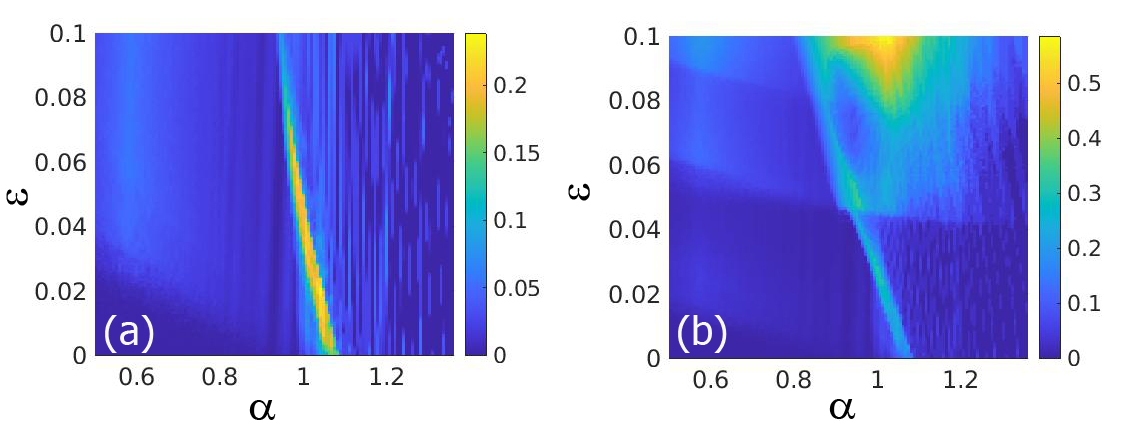}
   \caption{Escape parameter set $(\epsilon- \alpha)$,   for $F=0.2$ and $\phi=0.7\pi$ (a) and $\phi=\pi$ (b). In all panels, the color bar shows the ratio over $1$ of the trajectories that do not leave the potential well by varying in every point the initial conditions in the region $[x_0]\times [y_0]=[-1.5,0]\times[-0.7,0.7]$.}
\label{fig:8}
\end{figure}
\FloatBarrier

\begin{figure}[htbp]
  \centering
   \includegraphics[width=16cm,clip=true]{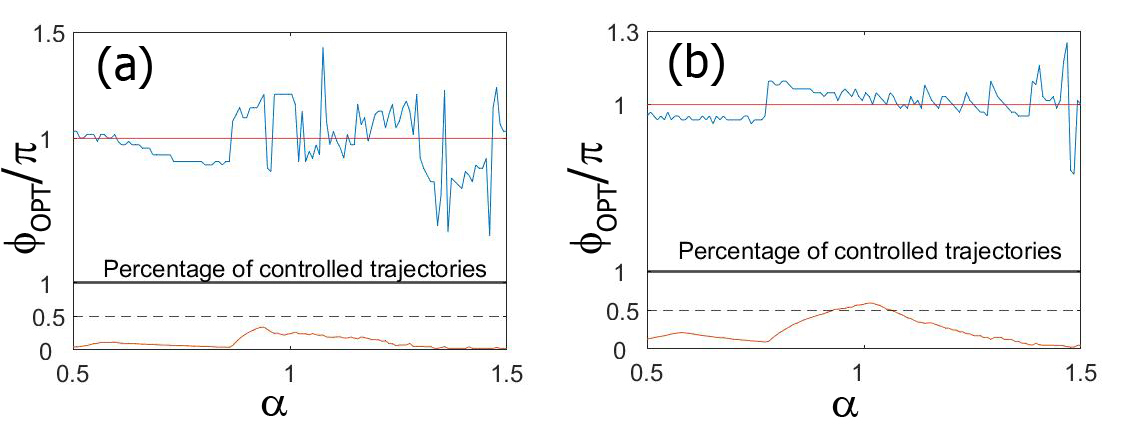}
   \caption{The optimal $\phi$ as a function of $\alpha$ (blue curve), for $F=0.2$ and $\epsilon=0.06$ (a) and $\epsilon=0.1$ (b). The panels show the $\phi$ value for which the ratio of the controlled trajectories is maximum for the initial conditions in the region $[x_0]\times [y_0]=[-1.5,0]\times[-0.7,0.7]$. The red solid line shows the value of $\pi$. The red curve, beneath the blue one, represents the percentage normalized to $1$ of the controlled trajectory for the represented $\phi_{OPT}$ along the $\alpha$ axis.}
\label{fig:9}
\end{figure}
\FloatBarrier

\section{Effect of the fractional parameter when the phase control parameters change} \label{sec:phi}

The effect of the fractional parameter in function of the phase $\phi$ is evident in Fig.~\ref{fig:4}. Here, we show in Fig.~\ref{fig:4}(a) and Fig.~\ref{fig:4}(b) the attractive region (the yellow points) and the escape times for $F=0.2$ and  $\epsilon=0.06$. Then,  we show in Fig.~\ref{fig:4}(c) and ~\ref{fig:4}(d) the attractive region and the escape times for $F=0.2$ and  $\epsilon=0.1$ as a function of the phase $\phi$ and the fractional parameter $\alpha$. We have denoted a color code bar to indicate the escape time of every trajectory as a function of the corresponding values of $\alpha$ and $\phi$. It is important to remark that the attractive regions give us at first glance the parameter values for which the particles remain bounded inside the well. On the other hand, the escape times show the parameters values for which we can modify the system dynamics to obtain longer or smaller time for the trajectory to escape. Of course, both are related since the region of maximum time delay are the yellow attractive region. Therefore, it can be useful to see them separately.
We have experienced along all the numerical simulations that, generally, the higher the intensity of the parameter $\epsilon$, the larger the area of the attractive region. In fact, in Fig.~\ref{fig:4}(a) the controlled trajectories are $3.5\%$ of the total, while in Fig.~\ref{fig:4}(c) we can control the $7.8\%$. Also, the escape times are strongly related with the attractive region topology. Similar behaviors are reported in Fig.~\ref{fig:5} where the attractive region and the escape times gradient plot are depicted in the parameter set $(\epsilon- \alpha)$.
The range of $\epsilon$ values has been selected to guarantee that the modulation term is smaller than the amplitude of the external periodic forcing. This ensures that the control is obtained through minimizing the impact of perturbations on the system and optimizing the energy used to bound the trajectories. We mean $\epsilon \in(0,0.1]$ which is smaller than $F=0.2$. So, in Fig.~\ref{fig:5}(a) and in Fig.~\ref{fig:5}(b) we have set $\phi=0.7\pi$ and in Fig.~\ref{fig:5}(c) and Fig.~\ref{fig:5}(d) $\phi=\pi$. The first $\phi$ value has been chosen because it is a value of $\phi$ that shows lots of escaping trajectories in Fig.~\ref{fig:4}, but also intersects briefly the attractive region. The second one is the value of $\phi$ that cuts the attractive region in the middle. The difference in controlled trajectories is evident. In fact, in Fig.~\ref{fig:5}(a) they are the $3.5\%$ of the total, while in Fig.~\ref{fig:5}(c) they are $13.8\%$.

The proposed figures express the deep impact of the $\alpha$ parameter on the phase control method. In fact, on the border of the attractive region a small change of the parameter can be pivotal to decide if the trajectory is controlled inside the well or not.

\section{Control of escapes on the initial conditions}\label{sec:alpha}

The importance of the fractional parameter on the phase control is well described by the figures presented until now. Nevertheless, to deepen the understanding of its impact, we decided to carry on the study by changing the initial conditions in order to complete the previous study carried out in the last sections. Therefore, we analyze the basins of attraction in the space of initial conditions. We define a basin of attraction as the set of initial conditions that lead to a certain attractor or fixed point. In our case, our attractors are the infinity (when the particle escapes) and the region inside the potential well (when the particles do not escape). For this purpose, we have depicted in Fig.~\ref{fig:6}, the basin of attractions in the left column and the escape time in the right column in the initial conditions space $[-1.5,0]\times[-0.7,0.7]$. Here, we have fixed $\epsilon=0.1$ and $F=0.2$ in Fig.~\ref{fig:6}, where the phase is $\phi=\pi$ in the first two panels and $\phi=0.7\pi$ in the second two panels. We indicate the percentages of the number of the controlled trajectories (denoted in yellow) thinking similarly to the framework called \textit{dynamical integrity} \cite{Thompson,Lenci}. This can be defined as the analysis that can provide valuable information about the expected dynamics under realistic conditions. It permits to safely operate a system with the desired behavior, depending on expected disturbances (for us the modulation term and the periodic forcing).  We can see the impact of the phase $\phi$ in the dynamic of the system when the initial conditions change. Therefore, we have decided to calculate how many trajectories can be controlled if the initial conditions change in the region stated above for different values of $\phi$ and $\alpha$. In Fig.~\ref{fig:7}, we plot the \textit{Escape parameter set}, that shows the percentage of controlled trajectories with initial conditions $[x_0\times y_0]=[-1.5,0]\times[-0.7,0.7]$, defined in a grid of $40\times40$, in the parameter set $(\phi- \alpha)$. Here, the percentage values are normalized to $1$ for a better display of the results. The parameters values are $F=0.2$ in both panels and $\epsilon=0.06$ in Fig.~\ref{fig:7}(a) and $\epsilon=0.1$ in Fig.~\ref{fig:7}(b). It is possible to spot the difference in the percentage values between the two figures due to the different amplitude of the modulation term. More important is to stress out the complexity of the effect of the fractional parameter on the escape ratio of the particles.
For a better understanding of the importance of the fractional parameter, we have carried out in Fig.~\ref{fig:8} a similar analysis in the parameter set $(\epsilon-\alpha)$. In Fig.~\ref{fig:8}(a) we have fixed $\phi=0.7\pi$ and in Fig.~\ref{fig:8}(b) $\phi=\pi$. It is possible to appreciate the difference in the percentage value between the two figures and how the effect of the interaction between $\epsilon$  and $\alpha$ is not simple. In fact, although the numerical simulations show that, generally, the trend should be the bigger the value of $\epsilon$ the bigger the percentage of the controlled trajectories. Notice that locally it is not always respected and we can observe there some fluctuations. Moreover, Fig.~\ref{fig:8}(a) shows a different trend and we can see that fact very clearly.

Finally, from the data of Fig.~\ref{fig:7} we extrapolate the optimal $\phi$ values that control the larger number of trajectories in function of the fractional parameter. These values are depicted in Fig.~\ref{fig:9}. Here, it is possible to appreciate a fluctuation around $\phi=\pi$ in both cases $\epsilon=0.06$ and $\epsilon=0.1$. In the case of $\epsilon=0.06$, where the control amplitude is smaller, the average value of the optimal phase is $\overline{\phi_{op}}=3.0913$, with a standard deviation of $\sigma=0.45$. On then other hand, for the case of $\epsilon=0.1$, the average value of the optimal phase is $\overline{\phi_{op}}=3.1479$, with a value of $\sigma=0.21$.
For $\epsilon=0.06$, the control amplitude is not so large and, therefore, the fluctuations around $\phi=\pi$ are more visible although the general trend is quite close to $\pi$, as corroborated by the value of the $\sigma$. In the case of $\epsilon=0.1$, in which the control amplitude is larger than the previous one, the fluctuations around $\phi=\pi$ are not so visible as in the previous case since the forcing amplitude is predominant in the effects between $\epsilon$ and $\alpha$. Again, the value of $\sigma$ confirms the analysis. It means that, in this last situation, the role of the fractional parameter $\alpha$ is negligible on the optimal value of $\phi$ and we mainly recover the non-fractional situation in which $\phi=\pi \simeq \phi_{op}$. This means that the smaller the $\epsilon$ value the stronger the influence of the fractional parameter on the optimal value of the phase $\phi$ and consequently on the control of the escapes. Nevertheless, we notice that the percentage of controlled trajectories differs significantly for each optimal phase value in the two cases. As illustrated by the red curves in Fig.~\ref{fig:9}, when $\epsilon=0.06$, the maximum number of trajectories that the optimal $\phi$ can control is in percentage less then $50\%$, while it is higher in the other case. Then, for the sake of clarity, we can integrate the $\phi_{OPT}$ curve along the $\alpha$ axis and calculate the percentage
\begin{equation}
    n=\frac{1}{N}\int_{\alpha_m}^{\alpha_M}\phi_{OPT}(\alpha)d\alpha,
\end{equation}
where $\alpha_m=0.5$, $\alpha_M=1.5$ and $N$ is the number of launched trajectories for each $\alpha$ value. So, in the first case, using the optimal $\phi$ values, we are able to control the $11.5\%$ of all the calculated trajectories, while in the second case we can control the $24.7\%$. Moreover, it is essential to highlight that the red curves vary significantly in function of the $\alpha$ parameter. This means that the proportion of the bounded trajectories is directly related to the value of $\alpha$. Therefore, this study underscores the impact of the fractional parameter on the percentage of controlled trajectories, even when utilizing the values of $\phi=\phi_{OPT}$.

To summarize, the main findings of this section are as follows: the phase control technique demonstrates robustness when varying the fractional parameter value, with the optimal $\phi$ value consistently oscillating around $\phi_{OPT}=\pi$. Conversely, altering $\alpha$ significantly affects the percentage of controlled trajectories when initial conditions vary.

\section{Conclusions and discussion}\label{sec:conclusions}
By using the Helmholtz oscillator with a fractional damping, we have analyzed the effects of a parametric control with a phase difference $\phi$ between the main forcing and the control. Afterwards, we have studied the influence of the fractional parameter $\alpha$ on the control. In the situation in which all particles are escaping from the potential well, $F=0.2$, in which $\alpha=1$ (non-fractional case), we have found the optimal values of the phase $\phi$ for different values of the fractional parameter. This optimal value varies with $\alpha$ but its average is close to the ones corresponding to the non-fractional case, $\phi\approx \pi$. The fluctuations of the optimal value of $\phi$ with respect the variation of $\alpha$ are illustrated in detail and they are due to the decrease or increase or the damping term. Although the fractional parameter may not significantly influence the main optimal value of $\phi$, it does play a crucial role in determining the proportion of controlled trajectories, even when utilizing an optimal value of $\phi=\phi_{OPT}$. Finally, and since the optimal value of the phase controlling the escapes oscillates around the value in the non-fractional case, we conjecture that the phase control technique is a robust way to control the escapes in an open dynamical systems.

\section{Acknowledgment}

This work has been supported by the Spanish State Research Agency (AEI) and the European Regional Development Fund (ERDF, EU) under Project No.~PID2019-105554GB-I0 (MCIN/AEI/10.13039/501100011033).

\end{document}